\documentclass[12pt]{article}
\usepackage{epsfig}
\usepackage{graphicx}
\usepackage{wrapfig}
\usepackage{boxedminipage}
\usepackage{setspace}
\usepackage{bm}
\usepackage{amsmath}
\usepackage{yfonts}
\usepackage{subfigure}
\usepackage{dcolumn}
\usepackage{bm}
\def\be{\begin{eqnarray}}\def\ba{\begin{eqnarray}}
\def\ee{\end{eqnarray}}\def\ea{\end{eqnarray}}
\def\ben{\begin{enumerate}}\def\bitem{\begin{itemize}}
\def\een{\end{enumerate}}\def\eitem{\end{itemize}}

\def\roughly#1{\mathrel{\raise.3ex\hbox{$#1$\kern-.75em%
\lower1ex\hbox{$\sim$}}}}

\def\A0{A_0}
\def\bq{\begin{equation}}
\def\eq{\end{equation}}
\def\la{\langle}\def\ra{\rangle}
\def\K0{K^0}

\begin{document}
\begin{titlepage}
\begin{center}

 \vskip 1.5cm

{\Large \bf Consequences of the partial restoration of chiral symmetry in AdS/QCD}
\vskip 1. cm
  {\large Youngman Kim$^{a}$ and  Hyun Kyu Lee$^{b}$ }

\vskip 0.5cm

(a)~{\it School of Physics, Korea Institute for Advanced Study,
Seoul 130-722, Korea}

(b)~{\it  Department of Physics and BK21 Division of Advanced
Research \\ and Education in Physics, Hanyang University, Seoul
133-791, Korea}

\end{center}

\centerline{(\today) }
\vskip 1cm
\vspace{1.0cm plus 0.5cm minus 0.5cm}

\begin{abstract}
Chiral symmetry is an essential concept in  understanding
QCD at low energy. We treat the chiral condensate, which measures the
spontaneous breaking of chiral symmetry, as a free parameter to
investigate the effect of partially restored chiral symmetry on
the physical quantities in the frame work of an AdS/QCD model. We
observe an interesting scaling behavior among the nucleon mass,
pion decay constant and chiral condensate.  We  propose a
phenomenological way to introduce the temperature dependence of a
physical quantity in the AdS/QCD model with the thermal AdS
metric.

\end{abstract}
\end{titlepage}

\section{Introduction}

Based on the AdS/CFT~\cite{adscft},  many successful attempts have been made to construct
a holographic model of QCD, in both the bottom-up~\cite{EKSS,PR}
 and the top-down\cite{sakai-sugimoto} approaches.
 Baryons are also introduced into both of the
 approaches~\cite{Baryon}.

  Since it has been generally expected that the physical properties of
 hadrons undergo a significant change in the finite
 temperature and/or density environment,
 finite temperature extension of the approaches have been of great
 interest. One of the interesting approaches is to use  the AdS black hole.  However, according to the
 Hawking-Page transition analysis
done in  \cite{Herzog},  the AdS black hole
is unstable at low temperature and the thermal AdS metric is
energetically favored in the confining phase~\cite{Herzog}. In
general, the thermal AdS background will not render any
temperature dependence of any physical quantities in confined phase, since the metric, thermal AdS, is
not globally modified compared to those at zero temperature, AdS.
It is shown in~\cite{thQCD} that
 in low-temperature confined phase, the properties of hadrons show
 no significant changes compared to zero temperature.
 This means that, in the light of the Hawking-Page
transition, the AdS/QCD model may not be of much
use, when it comes to the temperature dependence of physical quantities
such as meson and baryon masses. As long as we are taking $N_c\rightarrow \infty$ limit,
the observation of the Hawking-Page transition analysis is consistent with
large $N_c$ QCD~\cite{largeNQCD}. In real world, however,
the properties of hadrons, in confined phase,
are modified at finite temperature, see~\cite{GL89, PNJL, GL87, BK06, LQt} for examples.
In the light of the
Hawking-Page, we could think of such a temperature dependence as a
consequence of large $N_c$ corrections in an AdS/QCD approach. We
note here that up to now, however, such large $N_c$ corrections
have  not been successfully included in AdS/QCD \cite{HY}. In this
work we propose a simple  way to introduce a temperature
dependence  through the chiral condensate in the frame work of the hard
wall model~\cite{EKSS, PR}.

The chiral symmetry of QCD has been playing an important role in
hadron physics. A pertinent order parameter for the symmetry is
the chiral condensate $\la \bar qq \ra$.
There has been any amount of research on  the chiral
condensate at finite temperature and density
 in the framework of QCD effective theories or models.
 One of the interesting questions regarding the chiral condensate is: what is the effect of
 the partial restoration of the chiral symmetry on the physical quantities such as meson and baryon masses?

 In a bottom-up AdS/QCD model, the chiral condensate is
encoded in a 5D profile of a scalar field $X$ that couples to
quark bilinear $\bar qq$ at the boundary of AdS$_5$. The
background geometry of the hard wall model~\cite{EKSS,PR}  is
defined as a slice of anti-de Sitter (AdS) metric,
\begin{equation}
ds^2 = \frac{1}{z^2}\left( \eta_{\mu\nu} dx^{\mu} dx ^{\nu} -dz^2 \right),~~~z_0\leq z \leq z_m\, ,
\end{equation}
where $z_0\rightarrow 0$. Here $z_0$ is the UV-cutoff and
 $z_m$ for the IR-cutoff.
In the hard wall mode~\cite{EKSS,PR} the vacuum expectation of the scalar
$X_0=\la X(x,z)\ra$ is given by  $X_0=c_1z +c_2z^3$, where $c_1$
and $c_2$ are integration constants to be fixed by boundary
conditions. According to an AdS/CFT dictionary, $c_1$ is identified
with current quark mass, $c_1\sim m_q$, and $c_2$ is interpreted as the chiral
condensate, $c_2\sim$ $\la\bar qq\ra$.
In this work we will set $c_1=0$, no explicit chiral
symmetry breaking, and take $c_2$ as a free parameter of the model.
In the hard wall model, the correspondence $c_2\sim$ $\la\bar qq\ra$ is
realized by imposing the IR boundary condition:
$X_0(z_m)/z_m^3\sim \la\bar q q\ra$~\cite{PR}.
We vary the
value of $c_2$ from a finite value to zero to mimic a chiral symmetry restoration in a specific
environment such as QCD at finite temperature.

Our primary goal in this study is to observe how physical
quantities depend on the change of the chiral condensate, which is, in turn,
 supposed to address the consequences of the (partial)
restoration of chiral symmetry  to the physical quantities of QCD
effective theories at low energy. One of the interesting
observations as a consequence of chiral symmetry restoration is
the scaling behavior of the hadron properties  at finite
temperature and/or density. For examples, NJL model gives
~\cite{NJLs, BR96} \ba
\frac{m_N^\star}{m_N}\simeq\frac{m_\sigma^\star}{m_\sigma}\simeq\frac{\la\bar
qq\ra^\star}{\la\bar qq \ra}\, ,\label{Nambuscaling} \ea and BR
scaling ~\cite{BR96, BRs} reads \ba
\frac{m_N^\star}{m_N}\simeq\frac{m_\sigma^\star}{m_\sigma}\simeq\frac{m_\rho^\star}{m_\rho}\simeq\frac{m_\omega^\star}{m_\omega}
\simeq \frac{f_\pi^\star}{f_\pi}.\label{BRscaling} \ea Here
$\star$ is for temperature/density dependent quantities.

As described in \cite{EKSS, PR}, the chiral condensate in the hard
wall model is an integration constant to be fixed by the IR
boundary condition: $X_0(z_m)/z_m^3=c_2\sim \la\bar q q\ra$ with
$m_q=0$.  It has been  known ~\cite{GL89, PNJL, GL87, BK06, LQt}
that the value of the chiral condensate changes with temperature.
  Therefore we can define
a hard wall model at finite temperature with the following IR
boundary condition: $X_0(z_m)/z_m^3=c_2\sim \la\bar q q\ra^\star$.
Here $ \la\bar q q\ra^\star$ denotes the temperature dependent
chiral condensate. This is a proposal in this work  as   a simple
way to introduce a temperature dependence in the hard wall model
by imposing a IR boundary condition defined at finite temperature.
We cannot, however, determine the temperature dependence of the
chiral condensate in a self-consistent way within the hard wall
model. Therefore, as done in zero temperature
case~\cite{EKSS, PR}, we have to consider $\la \bar qq\ra^\star$
as an input,
 and take the temperature dependence of the chiral condensate from a QCD effective theory or lattice QCD study.
This is a limitation of the present study. Once the temperature
dependence of the chiral condensate is given, however, we can
easily obtain the temperature dependence of the other physical
quantities such as the nucleon mass and the pion decay constant.

\section{Scalings of physical quantities in the hard wall model}
The action of the model developed in~\cite{EKSS, PR} is, adopting the convention in~\cite{EKSS},
\begin{eqnarray}
 S_5&=&\int d^4 x \int dz \mathcal{L}_5 \nonumber \\
 &=& \int d^4 x \int dz \sqrt{g} ~{\rm Tr} [
-\frac{1}{4g_5^2}(F_L^2+F_R^2)\nonumber
\\ && \quad\quad\quad\quad+|DX |^2 +3|X^2|~]\label{S5}
\end{eqnarray}
where $D_\mu X = \partial_\mu X - i A_{L\mu} X+ i X A_{R\mu}$ and
$A_{L,R}=A^a_{L,R} t^a$ with ${\rm Tr}(t^a t^b)=\frac{1}{2}
\delta^{ab}$. The bulk scalar field is defined by  $X = X_0
e^{2i\pi^at^a}$, where $X_0 \equiv \langle X \rangle$. The
background metric of the model  is a slice of AdS metric,
\begin{equation}
ds^2 = \frac{1}{z^2}\left( \eta_{\mu\nu} dx^{\mu} dx ^{\nu} -dz^2 \right),~~~z_0\leq z \leq z_m\, ,
\end{equation}
where $z_0\rightarrow 0 $.
 Here $g_5$ is the 5D gauge coupling, $g_5^2=12\pi^2/N_c$, $z_0$ is the UV-cutoff and
 $z_m$ is for the
IR-cutoff. In~\cite{EKSS, PR}, the IR-cutoff $z_m$ is fixed by the
$\rho$-meson mass: $1/z_m \simeq 320~{\rm MeV}$. The vector- and
axial-vector mesons are  defined by
\begin{eqnarray}
&&V_\mu = \frac{1}{2}(A_L+A_R) \nonumber \\
&&A_\mu = \frac{1}{2}(A_L-A_R)\, .
\end{eqnarray}
To define our scaling factor $\sigma$, we solve the equation of
motion for $X_0$,
\ba \left[ \partial_z^2 -\frac{3}{z}\partial_z
+\frac{3}{z^2} \right]X_0=0,~~ X_0=c_1z +c_2z^3\, , \label{EoMX}
\ea
where $c_1$
and $c_2$ are integration constants. We define $v(z)\equiv 2 X_0$.
An AdS/CFT dictionary dictates
that $c_1$ is nothing but the source term, current quark mass $m_q$,
and $c_2$ should be interpreted as the chiral condensate, the order
parameter of chiral symmetry breaking/restoration. In this work, we
take  $m_q=0$. Then, we have
\ba
c_1=0,~~~c_2=\frac{1}{2}\sigma \, ,\nonumber
\ea
 where $\sigma$ is a free parameter of the model.
  Note that $\sigma=\sigma_0\approx (0.33~{\rm GeV})^3$ is compatible with phenomenology~\cite{EKSS}.
 We scale down $\sigma$ from $\sigma_0$ to zero.

\begin{figure}[!ht]
\begin{center}
{\includegraphics[angle=0,
width=0.53\textwidth]{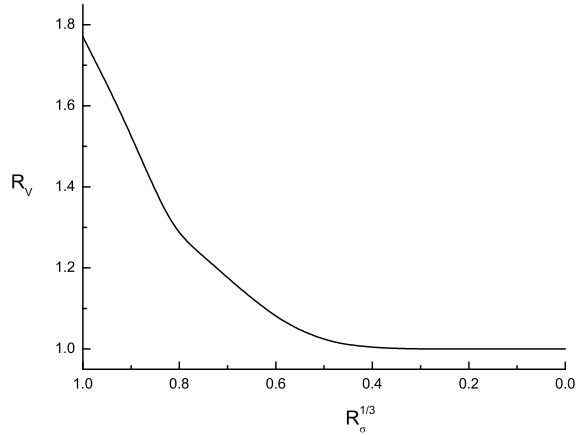}}
\caption{ \small  The scaling of $a_1$ mass normalized to $m_\rho$ as a function of $\sigma$.
 Here $R_V\equiv m_{a_1}(\sigma)/m_\rho$  and $R_\sigma^{1/3}=(\sigma/\sigma_0)^{1/3}$. } \label{mar}
\end{center}
\end{figure}

We first consider the scaling of vector and axial-vector meson
masses as a function of $\sigma$. The relevant equations of motions  for
the transverse component of the vector and axial-vector bulk fields
are, after the Kaluza-Klein reduction of the bulk field, $V_\mu
(x,z)= \Sigma_n f_n^V(z) V_\mu^{(n)}(x)$,
 \ba
&&\left [ \partial_z^2 -\frac{1}{z}\partial_z +m_n^2  \right ]f_n^V(z)=0\, ,\label{EoMV}\\
&&\left [ \partial_z^2 -\frac{1}{z}\partial_z
+m_n^2-g_5^2\frac{v^2}{z^2}  \right ]f_n^A(z)=0\label{EoMA}\, , \ea
where $v(z)=\sigma z^3$. As in \cite{PR}, we impose the following
boundary conditions: $f_n^{V,A}(z_0)=0,~\partial_z f_n^{V,A}(z_m)=0
$. Since the equation of motion for the vector is blind to $v(z)$,
 due to $D_\mu v=\partial_\mu v-2i vA_\mu$,
the 4D mass of the vector meson $m_n$ such as $\rho$-meson mass will
not scale with $\sigma$. While, the mass of axial-vector mesons will
change with varying $\sigma$. We plot the mass of the lowest lying
vector and axial vector mesons, $\rho$ and $a_1$, in Fig.~\ref{mar}.
As in Fig.~\ref{mar}, we have $m_{a_1}\approx m_\rho$ at
$R_\sigma^{1/3} (\equiv \sigma^{1/3}/\sigma_0^{1/3})\approx 0.4$.
This means that the role of the term with $v^2$ in Eq. (\ref{EoMA})
becomes negligible when $R_\sigma^{1/3}\le 0.4$.

\begin{figure}[!ht]
\begin{center}
\subfigure[] {\includegraphics[angle=0,
width=0.45\textwidth]{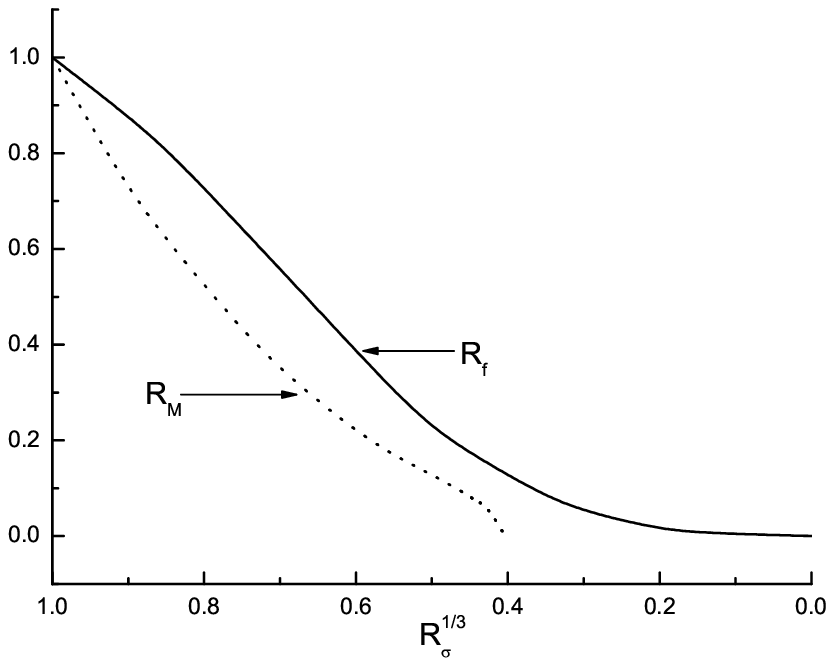} \label{mnf}}
\subfigure[] {\includegraphics[angle=0,
width=0.45\textwidth]{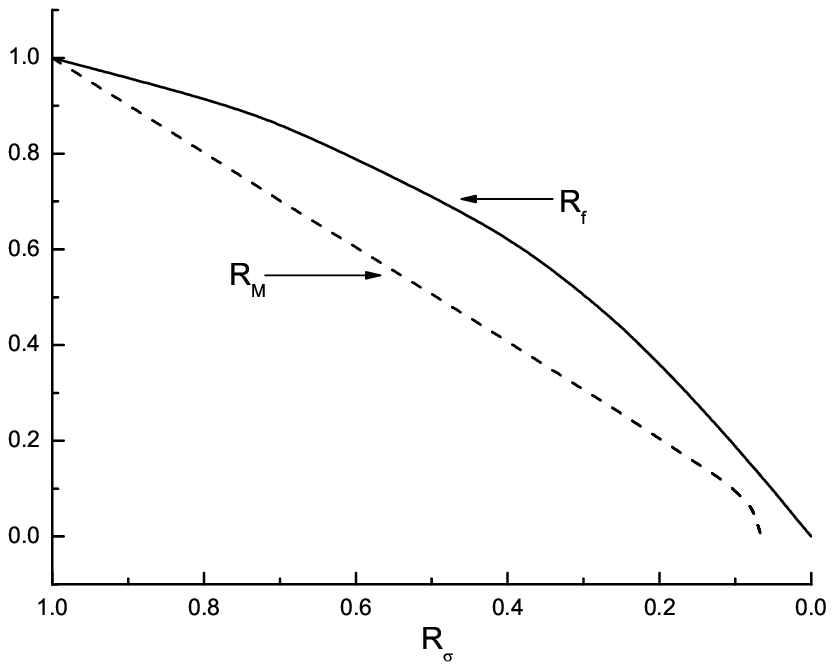} \label{mnf3}}
\caption{\small  The scaling of the nucleon mass and the pion decay constant as a function of (a)
$R_\sigma^{1/3}$, (b) $R_\sigma\equiv \sigma/\sigma_0$.
 Here $R_M\equiv m_{N}(\sigma)/m_N(\sigma_0)$  and $R_f\equiv f_\pi(\sigma)/f_\pi(\sigma_0)$. }\label{mnft}
\end{center}
\end{figure}

Now we discuss the pion decay constant.
The pion decay constant is defined by~\cite{EKSS}
\ba
f_\pi^2=-\frac{1}{g_5^2} \frac{\partial_z A(0,z)}{z}|_{z=z_0}\, ,\label{fpD}
\ea
where $A(0,z)$ is the solution of the following equation,
\ba
\left [ \partial_z^2 -\frac{1}{z}\partial_z -2\frac{v^2}{z^2}  \right ]A(0,z)=0\, .\label{fpA}
\ea
The scaling behavior of $f_\pi$ is shown in Fig.~\ref{mnft} together with that of nucleon.

Finally, we delve into the scaling behavior of the nucleon mass with $\sigma$.
The model~\cite{EKSS, PR} is extended to include baryons in Ref.~\cite{Ho}.
 The AdS/QCD model of spin ${1\over2}$, isospin ${1\over 2}$ baryons is
given by the action, referring to ~\cite{Ho} for details,
\be
S_{\rm kin}& =& \int dz \int dx^4 \sqrt{G_5}\,\left[i \bar N_1 \Gamma^M D_M N_1+i \bar N_2 \Gamma^M D_M N_2
-{5\over 2}\bar N_1 N_1
+{5\over 2}\bar N_2 N_2\right]\quad,\nonumber\\
S_m &=& \int dz \int dx^4 \sqrt{G_5}\,\left[ -g \bar N_1 X N_2 -g
\bar N_2 X^\dagger N_1\right]\quad,\label{action} \ee where the
covariant derivatives for $N_1$ and $N_2$ include the gauge group
$SU(2)_L \times SU(2)_R$ as well as the metric connection, and a
single parameter $g$ should be fixed to reproduce the nucleon mass.
 By expanding $N_1$ and $N_2$ in terms of KK modes,
it is easy to find the mode equations that must be solved to find
the mass spectrum of 4D spin $1\over 2$ baryons. Writing
$N_1(x,z)=f_{1L}(z)B_L(x)+f_{1R}(z)B_R(x)$ and similarly for
$N_2(x,z)=f_{2L}(z)B_L(x)+f_{2R}(z)B_R(x)$, where $B_{L,R}$ are the
components of the 4D  spinor $B=(B_L, B_R)^T$ with mass $m_N$ to be
determined, we have
\begin{eqnarray}
&&\left(
\begin{array}{cc}
\partial_z -{\Delta \over z} & -{ \frac{gX_0}{z } } \\
-{ \frac{gX_0}{z} } & \partial_z -{4-\Delta \over z}
\end{array}
\right)
\left(
\begin{array}{c} f_{1L} \\ f_{2L}
\end{array}
\right)=-m_N
\left(
\begin{array}{c} f_{1R} \\ f_{2R}
\end{array}
\right)\,, \nonumber\\
&&\left(
\begin{array}{cc}
\partial_z -{4-\Delta \over z} &{ \frac{gX_0}{z} }\\
{ \frac{gX_0}{z} } & \partial_z -{\Delta \over z}
\end{array}
\right)
\left(
\begin{array}{c} f_{1R} \\ f_{2R}
\end{array}
\right)=m_N
\left(
\begin{array}{c} f_{1L} \\ f_{2L}
\end{array}
\right)
\label{kk_mode}
\end{eqnarray}
with the IR boundary condition
$f_{1R}(z_m)=f_{2L}(z_m)=0$~\cite{Ho}. Here $\Delta=9/2$, and $
X_0= \frac{1}{2}\sigma z^3$. The scaling behavior of $m_N$ is
shown in Fig.~\ref{mnft}. Note that as in the case of axial vector
meson mass, $a_1$, in Fig.~\ref{mar}, the role of $X_0$ in the baryon mode equation becomes
negligible at  $R_\sigma^{1/3}\approx 0.4$ and
the mass of nucleon is
almost zero at and after $R_\sigma^{1/3}\approx 0.4$. We recast
Fig.~\ref{mnf} in Fig. \ref{mnf3}, where $R_\sigma\equiv
\sigma/\sigma_0$ is used for the horizontal axis. {}From  Fig.
\ref{mnf3}, we can see the  the overall scaling behavior of
$\frac{f_\pi(\sigma)}{f_\pi(\sigma_0)}$ is not much different from
$\frac{\sigma}{\sigma_0}$.  A salient feature of the hard wall
model is that the vector meson mass, for example $\rho$-meson
mass, is independent of the chiral condensate as shown in
Eq.~(\ref{EoMV}), and therefore the mass of $\rho$ is blind to the
restoration of chiral symmetry in the present approach. Finally,
we comment on the $\sigma$-dependence of the scalar meson mass. In
general, studying scalar excitations is not simple as they are
sensitive to the potential of $X$. If we follow Ref.\cite{PR2},
where a potential for $X$ is added on the IR, the mass of the
first scalar resonance is degenerate with the massless pion as
$\sigma\rightarrow 0$.

\section{Finite temperature as a boundary condition}

In the previous section, we obtain an interesting scaling
properties of physical quantities with respect to the chiral
condensate in the hard wall model.   The chiral condensate is
identified as the integration constant, $c_2$, fixed by a boundary
condition in Eq.(\ref{EoMX}).

In this section, we propose a phenomenological way to introduce
the temperature dependence of physical quantities in the hard
wall model by  imposing a boundary condition as a function of
temperature $T$. Then we have $c_2\sim \la\bar qq\ra^\star$, where
$\star$ is for temperature dependence. In general, in  bottom-up
AdS/QCD approach, however, there is no established way to
calculate the temperature dependence of $\sigma$. Therefore we
have to take $\sigma^\star$ as an input and borrow the temperature
dependence of $\sigma$ from a model calculation or lattice QCD
study. Although  it might be   a limitation of the present study,
once $\sigma^\star$ is given, we can easily obtain the temperature
dependence of the other hadronic parameters such as the nucleon
mass and the pion decay constant. Here we focus on the temperature
dependence of the pion decay constant.
\begin{figure}[!ht]
\begin{center}
\subfigure[] {\includegraphics[angle=0,
width=0.45\textwidth]{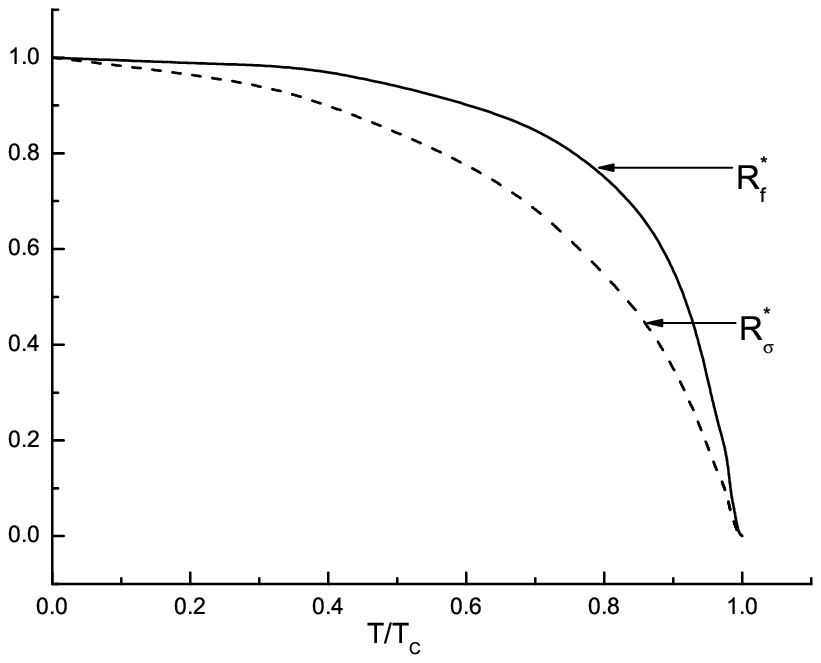} \label{fig:fp1}}
\subfigure[] {\includegraphics[angle=0,
width=0.45\textwidth]{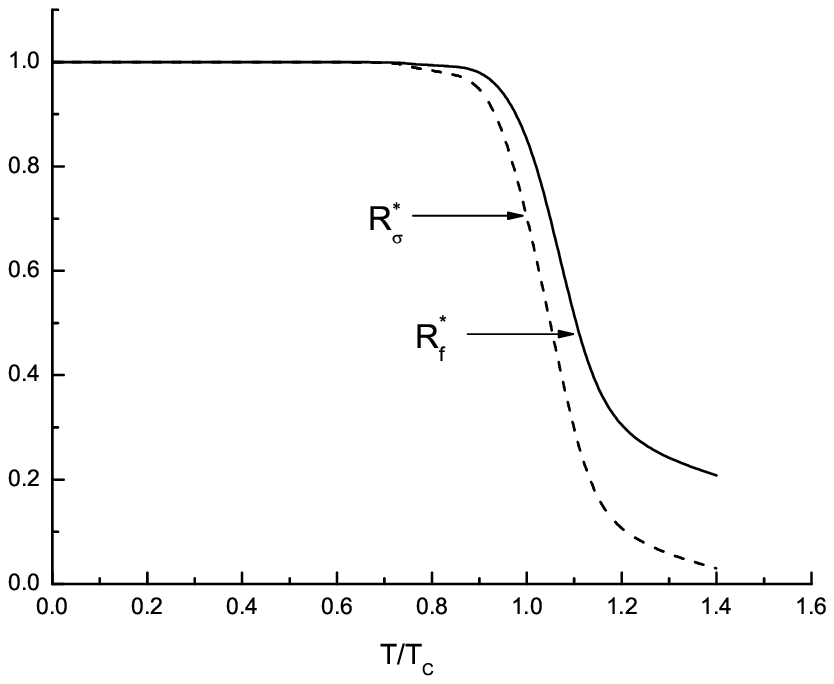} \label{fig:fp2}}
\caption{\label{fig:fpT}  \small   The temperature dependence of  the pion decay constant.
 Here the temperature dependent chiral
condensates are inputs, $R_\sigma^\star\equiv \sigma^\star/\sigma$ and
 $R_f^\star\equiv f_\pi^\star/f_\pi$:
(a) ${\rm R}_\sigma^\star$ taken from \cite{GL89} in the chiral limit,
 (b) ${\rm R}_\sigma^\star$ from ~\cite{PNJL}  with a finite current quark mass.
      }
\end{center}
\end{figure}
As an example, we take  $\sigma^\star/\sigma_0$ by  extrapolating
the temperature dependence of the chiral condensate obtained from  the
chiral Lagrangian~\cite{GL89} with two quark flavors in the chiral
limit, which is presumably valid at low temperature.  Then we can
 obtain the temperature dependence of $f_\pi$  using Eqs. (\ref{fpD}) and (\ref{fpA}) as shown in
Fig.~\ref{fig:fp1}.

In Fig. \ref{fig:fp2}, the temperature dependence of the pion decay
constant is calculated  by adopting  the temperature dependence of the
chiral condensate obtained in a generalized NJL model~\cite{PNJL} with a finite current quark mass.

Now, we compare the temperature dependence of $f_\pi$
calculated in this model with the one from chiral perturbation
theory~\cite{GL87} and from linear and non-linear sigma models
\cite{BK06}. We note here that at low temperature or at the leading order in temperature $T$,
both studies~\cite{GL87, BK06} give the same results on the chiral
condensate,
\ba
\frac{\sigma^\star}{\sigma_0} =1-\frac{T^2}{8f_\pi^2}\label{qC5}.
\ea
\begin{figure}
\begin{center}
{\includegraphics[angle=0,
width=0.53\textwidth]{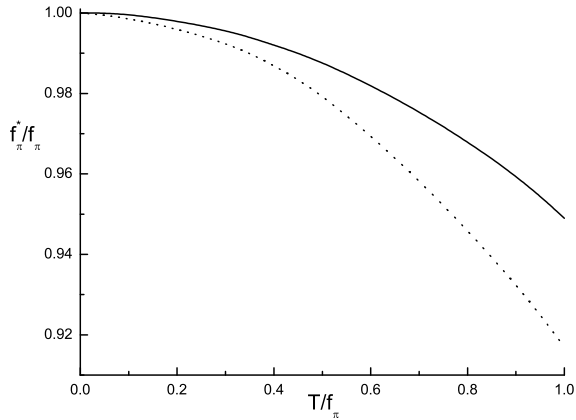}}
\caption{ \small  The temperature dependence of  the pion decay constant. The temperature dependent chiral
condensate is taken from~\cite{GL87, BK06}, Eq.~(\ref{qC5}). Solid line is for our result,
 and dashed line is from~\cite{GL87, BK06}, given in Eq.~(\ref{fp5}).
} \label{fpt}
\end{center}
\end{figure}
We take the temperature dependent chiral condensate $\sigma^\star/\sigma_0$ given in Eq. (\ref{qC5}) as an input
and calculate the temperature dependence of the pion decay constant using  Eqs. (\ref{fpD}) and (\ref{fpA}).
In Fig.~\ref{fpt}, we compare our $f_\pi^\star$ with that from~\cite{GL87, BK06}.
The temperature dependent pion decay constant obtained in ~\cite{GL87, BK06} at low temperature is
 given by
\ba
\frac{f_\pi^\star}{f_\pi} =1-\frac{T^2}{12f_\pi^2}\label{fp5}\, .
\ea
The observed temperature dependence of $f_\pi$ in our work is weaker than the one from~\cite{GL87, BK06}.
The temperature dependence obtained in the present work is only due to the temperature dependent chiral
condensate. In addition to this, we expect some temperature dependence due to large $N_c$ corrections and
from higher dimensional terms such as $F_L X F_R X^{\dagger}$ in the action of the hard wall model.

Finally, we remark that the temperature dependence obtained in the present study has some similarity with
that from QCD sum rule at finite temperature, for instance, see \cite{tQSR}.
In thermal QCD sum rule, the temperature dependence of condensates, {\it e.g.}, chiral condensate,
is taken from a model study or from lattice QCD, and it is conveyed to the (part of) temperature dependence of physical quantities such as hadron masses. Since the temperature dependent chiral condensate used in this work and the one adopted in the thermal QCD sum rule studies
 are different, we don't make a direct comparison of our result with that from \cite{tQSR}.
We note here that in  thermal QCD sum rule~\cite{tQSR}, the temperature dependence could have additional sources other than
the chiral condensate such as the temperature dependent Wilson coefficients or four-quark condensate.

\section{Summary}
We have used the hard wall model to study the scaling property of physical quantities as we scale down the chiral
condensate.
By varying the value of the chiral condensate $\sigma$,  we study how the other physical parameters scale with
$\sigma$, which is summarized in Fig.~\ref{mnft}.
We note here that the mass of $\rho$-meson is independent of the chiral condensate in the hard wall model~\cite{EKSS, PR}, and therefore
it is blind to the restoration of the chiral symmetry in the present work.

We have introduced the temperature into the AdS/QCD model through the chiral condensate
 as an IR boundary condition.
With a given temperature dependent chiral condensate $\sigma^\star$,
 we can easily predict the temperature dependence of other physical quantities such as  hadron masses or decay constants.
As an example, we have calculated the temperature dependence of the pion decay constant and compare our result with
that from chiral perturbation theory at low temperature.
We find that the temperature dependence of the pion decay constant predicted from the present study is
weak compared to that from chiral perturbation theory~\cite{GL87} and from linear and non-linear sigma models~\cite{BK06}.

\vskip 1cm
\noindent
{\large\bf Acknowledgments}\\
We thank Mannque Rho for useful comments. HKL was supported by the
Korea Science \& Engineering Foundation(KOSEF) grant funded by
Korea government(MOST)(No. R01-2006-000-10651-0).

\vskip 1cm


\end{document}